\documentclass{aip-cp}

\usepackage[numbers]{natbib}
\usepackage{rotating}
\usepackage{graphicx}
\usepackage{amsfonts}

  \def\fH{{\cal H}} \def\fK{{\cal K}}
\def\fE{{\cal E}} \def\fA{{\cal A}}

 \def\fU{{\cal U}}

\newcommand{\qed}{\hbox{\rule{6pt}{6pt}}}

\begin{document}

\title{Note on bounds for symmetric divergence measures}

\author[aff1]{S.Furuichi\corref{cor1}}
\author[aff2]{K.Yanagi}
\author[aff3]{K.Kuriyama}

\affil[aff1]{Nihon University}
\affil[aff2]{Josai  University}
\affil[aff3]{Yamaguchi University}
\corresp[cor1]{Corresponding author: furuichi@chs.nihon-u.ac.jp}

\maketitle

\begin{abstract}
In the paper \cite{Sas2015}, the tight bounds for symmetric divergence measures are derived by applying the results established in the paper \cite{Gil2006}.
In this article, we are going to report two kinds of extensions for the above results, namely classical $q$-extension and non-commutative(quantum) extension.
\end{abstract}

\section{INTRODUCTION}
In the paper \cite{Sas2015}, the tight bounds for symmetric divergence measures are derived by applying the results established in the paper \cite{Gil2006}.
 In the paper \cite{Sas2015}, the minimization problem for Bhattacharyya coefficient, 
Chernoff information, 
Jensen-Shannon divergence and Jeffrey's divergence under the constraint on total variation distance. 
In this article, we are going to report two kinds of extensions for the above results, namely classical $q$-extension and non-commutative(quantum) extension.
The parametric $q$-extension means that Tsallis entropy $H_q(X) \equiv \sum_{x} \frac{p(x)^q-p(x)}{1-q}$ \cite{Tsa1988} converges to Shannon entropy when $q \to 1$. Namely, all results with the parameter $q$ recover the usual (standard) Shannon's results when $q \to 1$.
We give here list of our extensions as follows.
\begin{itemize}
\item[(i)] The lower bound for Jensen-Shannon-Tsallis diverence is given by applying the results in \cite{Gil2006}.
\item[(ii)] The lower bound for Jeffrey-Tsallis divergence is given by applying the results in  \cite{Gil2006} and deriving $q$-Pinsker's inequality for $q \geq 1$. This implies new upper bounds of
$\sum_{u \in \fU} \vert p(u) -Q_{d,l}(u) \vert $.
\item[(iii)] The lower bound for quantum Chernoff information is given by the known relation between the trace distance and fidelity.
\item[(iv)] The lower bound for quantum Jeffrey divergence is given by applying the monotonicity (data processing inequality) of quantum $f$-divergence.
\end{itemize}


\section{$q$-EXTENDED CASES}
Here we review some quantities. The total variation distance between two probability distributions $P(x)$ and $Q(x)$ is defined by
$$
d_{TV}(P,Q) \equiv \frac{1}{2} \sum_x \vert P(x) -Q(x) \vert = \frac{1}{2} \vert\vert P-Q \vert \vert _1,
$$
where $\vert \vert \cdot \vert \vert_1$ represents $l_1$ norm.
The $f$-divergence introduced by Csisz\'{a}r in \cite{Csi1967} is defined by
$$
D_f(P\vert\vert Q) \equiv \sum_x Q(x) f\left( \frac{P(x)}{Q(x)}\right)
$$
where $f$ is convex function and $f(1)=0$. If we take $f(t) =-t \ln_q \frac{1}{t}$, where $\ln_q(x) \equiv \frac{x^{1-q}-1}{1-q}$ is $q$-logarithmic function defined for $x \geq 0$ and $q \neq 1$, then $f$-divergence is equal to the Tsallis relative entropy (Tsallis divergence) defined by (see e.g., \cite{FYK2004})
$$
D_q(P\vert\vert Q) \equiv -\sum_x P(x)\ln_q \frac{Q(x)}{P(x)}=\sum_x \frac{P(x)-P(x)^qQ(x)^{1-q}}{1-q}.
$$
In this section, we use the result established by Gilardoni in \cite{Gil2006} for the symmetric divergence.
\\

{\bf Theorem (Gilardoni, 2006 \cite{Gil2006})}
We suppose $D_f$ is symmetric divergence (which condition is known as $f(u) =u f(1/u) + c (u-1)$, $u \in (0,\infty)$ and $c$ is constant number) and $f:(0,\infty) \to \mathbb{R}$ with $f(1)=0$. Then we have
\[\mathop {\inf }\limits_{P,Q:{d_{TV}}\left( {P,Q} \right) = \varepsilon } {D_f}\left( {P\left\| Q \right.} \right) = \left( {1 - \varepsilon } \right)f\left( {\frac{{1 + \varepsilon }}{{1 - \varepsilon }}} \right) - 2f'\left( 1 \right)\varepsilon \]
\\

As corollaries of the above theorem, we obtain the following two propositions.
We define the Jensen-Shannon-Tsallis diverence as
\[\overline {{C_q}} \left( {P,Q} \right) \equiv {D_q}\left( {P\left\| {\frac{{P + Q}}{2}} \right.} \right) + {D_q}\left( {Q\left\| {\frac{{P + Q}}{2}} \right.} \right).\]
Then ${D_{{f_q}}}\left( {P\left\| Q \right.} \right) = \overline {{C_q}} \left( {P,Q} \right)$ with ${f_q}\left( t \right) =  - t{\ln _q}\frac{{t + 1}}{{2t}} - {\ln _q}\frac{{t + 1}}{2}$,  ${f_q}$ is convex, with ${f_q}\left( 1 \right) = 0$ and $\overline {{C_q}} \left( {P,Q} \right) = \overline {{C_q}} \left( {Q,P} \right)$. Thus we have the following proposition which is $q$-parametric extension of Proposition 3 in \cite{Sas2015}.
\\

{\bf Proposition 1}
\[\mathop {\min }\limits_{P,Q:{d_{TV}}\left( {P,Q} \right) = \varepsilon } \,\overline {{C_q}} \left( {P,Q} \right) =  - \left( {1 - \varepsilon } \right){\ln _q}\frac{1}{{1 - \varepsilon }} - \left( {1 + \varepsilon } \right){\ln _q}\frac{1}{{1 + \varepsilon }},\]
The equality is archived when $P = \left( {\frac{{1 - \varepsilon }}{2},\frac{{1 + \varepsilon }}{2}} \right),Q = \left( {\frac{{1 + \varepsilon }}{2},\frac{{1 - \varepsilon }}{2}} \right)$.
\\

We also define Jeffrey-Tsallis divergence as
\[{J_q}\left( {P,Q} \right) \equiv \frac{1}{2}\left\{ {{D_q}\left( {P\left\| Q \right.} \right) + {D_q}\left( {Q\left\| P \right.} \right)} \right\}.\]
Then ${D_{{f_q}}}\left( {P\left\| Q \right.} \right) = {J_q}\left( {P,Q} \right)$ with ${f_q}\left( t \right) = \frac{{\left( {{t^q} - 1} \right){{\ln }_q}t}}{2}$,
${f_q}$ is convex with ${f_q}\left( 1 \right) = 0$ and ${J_q}\left( {P,Q} \right) = {J_q}\left( {Q,P} \right)$.
Thus we have the following proposition which is $q$-parametric extension of Proposition 4 in \cite{Sas2015}.
\\

{\bf Proposition 2}
\[\mathop {\min }\limits_{P,Q:{d_{TV}}\left( {P,Q} \right) = \varepsilon } \,{J_q}\left( {P,Q} \right) =  - \frac{1}{2}\left\{ {\left( {1 + \varepsilon } \right){{\ln }_q}\frac{{1 - \varepsilon }}{{1 + \varepsilon }} + \left( {1 - \varepsilon } \right){{\ln }_q}\frac{{1 + \varepsilon }}{{1 - \varepsilon }}} \right\}.\]
The equality is archived when $P = \left( {\frac{{1 - \varepsilon }}{2},\frac{{1 + \varepsilon }}{2}} \right),Q = \left( {\frac{{1 + \varepsilon }}{2},\frac{{1 - \varepsilon }}{2}} \right)$.
\\

Here we are able to prove the following lemma, which may be named $q$-Pinsker's inequality.
\\

{\bf Lemma 1}
\[{D_q}\left( {P\left\| Q \right.} \right) \ge \frac{1}{2}{d_{TV}}{\left( {P,Q} \right)^2}\quad for\quad q \geq 1.\]

{\it Proof:}
The proof is easily done by the fact that 
$\log t \le \frac{{{t^r} - 1}}{r},\left( {t > 0,r > 0} \right)$ implies $- \log \frac{1}{t} \le  - {\ln _q}\frac{1}{t},\left( {t > 0,q > 1} \right)$, putting $r=q-1$. Thus we have
\[ - x{\ln _q}\frac{y}{x} - \left( {1 - x} \right){\ln _q}\frac{{1 - y}}{{1 - x}} \ge  - x\log \frac{y}{x} - \left( {1 - x} \right)\log \frac{{1 - y}}{{1 - x}} \ge 2{\left( {x - y} \right)^2}\]
for ${0 < x,y < 1,q \ge 1}$. Thus we have this lemma by data processing inequality.

\hfill \qed
\\

As remark, the above  $q$-Pinsker inequality does not hold for the case $0<q<1$, since we have counter-examples. Applying this lemma, we can prove the following proposition, which condition is same to the paper \cite{Sas2015}
except for the extended parameter $q$.
\\

{\bf Theorem 1}
Consider a memoryless stationary source with alphabet $\fU$ with probability distribution $P$ and assume that a uniquely decodable code with an alphabet size $d$. For $q \geq 1$, we have

\[\frac{1}{2}\sum\limits_{u \in \fU} {\left| {p\left( u \right) - {Q_{d,l}}\left( u \right)} \right|}  \le \min \left\{ {1,\sqrt {\frac{{{\Delta _{d,q}}{{\log }_e}d}}{2}} } \right\}.\]

Where ${\Delta _{d,q}} \equiv {\overline n _q} - {H_{d,q}}\left( \fU \right)$,${\overline n _q} \equiv -\frac{\left( c_{d,l}\right)^{q-1 }}{\log_e d} 
\sum\limits_{u \in \fU } {p{{\left( u \right)}^q}  \ln_q d^{-l\left( u \right)}  }$,${H_{d,q}}\left( \fU \right) \equiv  - \frac{1}{{{{\log }_e}d}}\sum\limits_{u \in \fU} {p{{\left( u \right)}^q}{{\ln }_q}p\left( u \right)}$,
${Q_{d,l}}\left( u \right) \equiv \frac{d^{-l(u)}}{{{c_{d,l}}}} $ and 
 ${c_{d,l}} \equiv \sum\limits_{u \in \fU} d^{-l(u)}.$

{\it Proof:}
We give the sketch of the proof of this proposition. Firstly $\sum\limits_{u \in \fU} {\left| {p\left( u \right) - {Q_{d,l}}\left( u \right)} \right|} \leq 2$ is trivial.
By Lemma 1, we have 
\[{D_q}\left( {P\left\| {{Q_{d,l}}} \right.} \right) \ge {D_q}\left( {\widehat P\left\| {\widehat {{Q_{d,l}}}} \right.} \right) \ge 2{\left( {P\left( A \right) - {Q_{d,l}}\left( A \right)} \right)^2} = 2{\left( {\frac{1}{2}{d_{TV}}\left( {P,{Q_{d,l}}} \right)} \right)^2} = \frac{1}{2}{\left( {\sum\limits_{u \in \fU} {\left| {p\left( u \right) - {Q_{d,l}}\left( u \right)} \right|} } \right)^2},\]
where $A \equiv \left\{ {x:P\left( x \right) > {Q_{d,l}}\left( x \right)} \right\}$, $Y \equiv \phi \left( X \right)$ and $\widehat P$ and $\widehat {{Q_{d,l}}}$ are distributions of new random variable $Y$.
By simple computations with formula $\ln_q\frac{y}{x} =x^{q-1}(\ln_q y-\ln_q x)$, we have
\begin{eqnarray*}
{D_q}\left( {P\left\| {{Q_{d,l}}} \right.} \right)&=&\sum_{u\in \fU} p(u)^q \left(\ln_q p(u) - \ln_q Q_{d,l}(u)\right) = \sum_{u\in \fU} p(u)^q \left(\ln_q p(u) - \ln_q\frac{d^{-l(u)}}{c_{d,l}}\right) \\
 &=& -\log_e d \cdot H_{d,q}(\fU) - \left( c_{d,l}\right)^{q-1} \sum_{u\in \fU} p(u)^q \left( \ln_q d^{-l(u)} -\ln_q c_{d,l}\right) \\
&=& -\log_e d \cdot H_{d,q}(\fU) +\log_e d \cdot {\overline n_q} - \ln_q \frac{1}{c_{d,l}} \sum_{u\in\fU} p(u)^q \leq \log_e d \cdot \Delta_{d,q}
\end{eqnarray*}
since the Kraft-McMillian inequality $c_{d,l} \leq 1$ was used.
Thus we have $\frac{1}{2}\left( \sum\limits_{u \in \fU} {\left| {p\left( u \right) - {Q_{d,l}}\left( u \right)} \right|}  \right)^2 \le {\log _e}d\cdot \,{\Delta _{d,q}}.$

\hfill \qed
\\

{\bf Remark 1}
This theorem is a parametric extension of the inequality (32) in the paper \cite{Sas2015} in the sense that the left hand side of our inequality contains the parameter $q \geq 1$. We also note that the condition $q \geq 1$ is corresponding to the result in our previous paper \cite{Fur2006}, so the condition $q \geq 1$ may not be so unnatural within our framework of this topic. 

In addition, we compare our upper bound with parameter $q \geq 1$ obtained in Theorem 1 and that obtained in the paper \cite{Sas2015}. Actually we give an example such that $\sqrt{ \frac{\Delta_{d,q} \log_e d}{2}  } \leq \sqrt{\frac{\Delta_{d,1}\log_e d}{2}}$, where $\Delta_{d,1}$ was used in the paper \cite{Sas2015} as $\Delta_d$. Consider the following information source
\[\fU = \left( \begin{array}{l}
\,\,{u_1},\,\,\,\,\,{u_2},\,\,\,\,\,{u_3}\\
0.5,\,\,\,0.3,\,\,\,0.2
\end{array} \right),\]
with $d=2$. Then we have the code $u_1\to \lq\lq 0",u_2 \to \lq\lq10",u_3 \to \lq\lq 110"$ by Shannon-Fano coding, so that $c_{d,l} =\frac{7}{8} <1$ since $l_1=1, l_2=2,l_3=3$. By numerical computations, we have $\sqrt{ \frac{\Delta_{2,1.5} \log_e 2}{2}  } \simeq 0.225793$ and  $\sqrt{\frac{\Delta_{2,1}\log_e 2}{2}} \simeq 0.272669$. This means there exists a code such that $\sqrt{ \frac{\Delta_{d,q} \log_e d}{2}  } \leq \sqrt{\frac{\Delta_{d,1}\log_e d}{2}}$, which shows our upper bound with the parameter $q \geq 1$ is tighter than the upper bound in the paper \cite{Sas2015}, in this example. We performed some numerical computations with a few information sources, then we could find the parameter $q \geq 1$ such that  $\sqrt{ \frac{\Delta_{d,q} \log_e d}{2}  } \leq \sqrt{\frac{\Delta_{d,1}\log_e d}{2}}$ for the case $c_{d,l} < 1$.
\\

 However, for the case $c_{d,l} = 1$ (e.g., Huffman code), the following proposition can be proven.
\\

{\bf Proposition 3}
Let $q \geq 1$ and $c_{d,l} = 1$. Then we have the relation $\Delta_{d,1} \leq \Delta_{d,q}$.

{\it Proof:}
We firstly prove the inequality $f_q(x,y) \geq 0$ for $q \geq 1, 0 < x,y\leq 1$,
where $f_q(x,y)  \equiv x(\log_e y -\log_e x) +x^q(\ln_q x-\ln_q y).$
Since $\frac{df_q(x,y)}{dy}=\frac{x^q}{y^q} \left( \frac{x^{1-q}}{y^{1-q}} -1\right)$,
if $x \leq y$, then $\frac{df_q(x,y)}{dy} \geq 0$ and if  $x \geq y$, then $\frac{df_q(x,y)}{dy} \leq 0$, thus we have $f_q(x,y) \geq f_q(x,x) =0$.
Putting $x=p(u)$ and $y=d^{-l(u)}$, taking summation on both sides by $u\in \fU$ and dividing the both sides by $\log_e d$, we have
$$
-\frac{1}{\log_e d} \sum_{u\in\fU} p(u)^q \ln_q d^{-l(u)}+\frac{1}{\log_e d} \sum_{u\in\fU} p(u)\log_e d^{-l(u)} -\frac{1}{\log_e d} \sum_{u\in\fU} p(u)\log_e p(u) +\frac{1}{\log_e d} \sum_{u\in\fU} p(u)^q \ln_q p(u) \geq 0.
$$ 
When $c_{d,l} =1$, we thus obtain the inequality $\Delta_{d,q} -\Delta_{d,1} = {\overline n_q} -{\overline n_1} +H_{d,1}(\fU) -H_{d,q}(\fU) \geq 0$, taking account that the usual average code length can be rewritten as $ {\overline n_1} = \sum_{u\in\fU} p(u) l(u)= -\frac{1}{\log_e d} \sum_{u\in \fU} p(u)\log_e d^{-l(u)}$.

\hfill \qed
\\

This proposition shows that for the special (but nontrivial) case $c_{d,l}=1$, the upper bound $\sqrt{ \frac{\Delta_{d,1}  \log_ed}{2}}$ given in (32) of the paper \cite{Sas2015} is always tighter than ours $\sqrt{ \frac{\Delta_{d,q}  \log_ed}{2}}$ (for $q \geq 1$) obtained in Theorem 1.

\section{NON-COMMUTATIVE CASES}
Let $\rho$ and $\sigma$ be density matrices (quantum states), which are positive semi-definite matrices and unit trace. Then the following quantities are well known in the field of quantum information or physics as trace distance and fidelity, respectively:
\[d\left( {\rho ,\sigma } \right) \equiv \frac{1}{2}Tr\left| {\rho  - \sigma } \right|, \quad F\left( {\rho ,\sigma } \right) \equiv Tr\left| {{\rho ^{1/2}}{\sigma ^{1/2}}} \right|,\]
Where $\vert A \vert = (A^*A)^{1/2}$.
Then we have the following propositions.
\\

{\bf Proposition 4}
For the trace distance and fidelity, we have the following relation:
 \[1 - d\left( {\rho ,\sigma } \right) \le F\left( {\rho ,\sigma } \right) \le \sqrt {1 - d{{\left( {\rho ,\sigma } \right)}^2}}. \]
\\

This relation is well known in the field of quantum information or quantum statistical physics, and this proposition is non-commutative extension of Proposition 1 in the paper \cite{Sas2015}.

By the easy calculations such as 
${C_Q}\left( {\rho ,\sigma } \right) \equiv  - \log \left( {\mathop {\min }\limits_{0 \le s \le 1} Tr\left[ {{\rho ^s}{\sigma ^{1 - s}}} \right]} \right) =  - \mathop {\min }\limits_{0 \le s \le 1} \left( {\log Tr\left[ {{\rho ^s}{\sigma ^{1 - s}}} \right]} \right)\, \ge  - \log Tr\left[ {{\rho ^{1/2}}{\sigma ^{1/2}}} \right] \ge  - \log Tr\left[ {\left| {{\rho ^{1/2}}{\sigma ^{1/2}}} \right|} \right] =  - \log F\left( {\rho ,\sigma } \right) \ge  - \frac{1}{2}\log \left( {1 - d{{\left( {\rho ,\sigma } \right)}^2}} \right)$,
we have the following proposition.
\\

{\bf Proposition 5}
For the quantum Chernoff information, we have
\[\mathop {\min }\limits_{\rho ,\sigma :d\left( {\rho ,\sigma } \right) = \varepsilon } {C_Q}\left( {\rho ,\sigma } \right) = \left\{ \begin{array}{l}
 - \frac{1}{2}\log \left( {1 - {\varepsilon ^2}} \right),\varepsilon  \in \left[ {0,1} \right)\\
\,\,\, + \infty ,\,\,\,\,\,\,\,\,\,\,\,\,\,\,\,\,\,\,\,\,\,\,\,\,\varepsilon  = 1
\end{array} \right.\]
\\

The above proposition is also non-commutative extension of Proposition 2 in the paper \cite{Sas2015}.

The quantum  Pinsker inequality on quantum relative entropy (divergence) and similar one are known (see e.g., \cite{Pet2004} and \cite{Lie2014}, respectively)
\[D\left( \rho \vert \sigma   \right) \equiv Tr[\rho(\log \rho -\log \sigma)]\ge \frac{1}{2}Tr{\left[ {\left| {\rho  - \sigma } \right|} \right]^2}\,\]
and 
\[D\left( \rho \vert \sigma  \right) \ge  - 2\log Tr\left[ {{\rho ^{1/2}}{\sigma ^{1/2}}} \right] \ge Tr{\left[ {{\rho ^{1/2}} - {\sigma ^{1/2}}} \right]^2}\]

To show our final result, we use the following well-known fact.
See \cite{Pet2004} for example.
\\

{\bf Lemma 2}
Let $\fE : B(\fH) \to B(\fK)$ be a state transformation. For an operator monotone decreasing function $f:\mathbb{R}^+ \to \mathbb{R}$, the monotonicity holds:
\[{D_f}\left( {\rho \left| \sigma  \right.} \right) \ge {D_f}\left( {\fE\left( \rho  \right)\left| {\fE\left( \sigma  \right)} \right.} \right)\]
where ${D_f}\left( {\rho \left| \sigma  \right.} \right) \equiv Tr\left[ {\rho f\left( \Delta  \right)\left( I \right)} \right]$ is the quantum $f$-divergence, with ${\Delta _{\sigma ,\rho }} \equiv \Delta  = LR$ is the relative modular operator such as $L\left( A \right) = \sigma A$ and $R\left( A \right) = A{\rho ^{ - 1}}$.
\\

{\bf Theorem 2}
The quantum Jeffrey divergence defined by
$J\left( {\rho \left| \sigma  \right.} \right) \equiv \frac{1}{2}\left\{ {D\left( {\rho \left| \sigma  \right.} \right) + D\left( {\sigma \left| \rho  \right.} \right)} \right\}$ has the following lower bound:
\[J\left( {\rho \left| \sigma  \right.} \right) \ge d\left( {\rho ,\sigma } \right)\log \left( {\frac{{1 + d\left( {\rho ,\sigma } \right)}}{{1 - d\left( {\rho ,\sigma } \right)}}} \right).\]

{\it Proof:}
By Lemma 2, Proposition 4 in the paper \cite{Sas2015} and ${\left\| {\rho  - \sigma } \right\|_1} = {\left\| {P - Q} \right\|_1}$(which will be shown in the end of proof), we have 
\[J\left( {\rho \left| \sigma  \right.} \right) \ge J\left( {P\left| Q \right.} \right) \ge {d_{TV}}\left( {P,Q} \right)\log \left( {\frac{{1 + {d_{TV}}\left( {P,Q} \right)}}{{1 - {d_{TV}}\left( {P,Q} \right)}}} \right) = d\left( {\rho ,\sigma } \right)\log \left( {\frac{{1 + d\left( {\rho ,\sigma } \right)}}{{1 - d\left( {\rho ,\sigma } \right)}}} \right).\]
Here we note that $f\left( t \right) = \frac{1}{2}\left( {t - 1} \right)\log t$ is operator convex which is equivalent to operator monotone decreasing and we have  ${D_{\frac{1}{2}\left( {t - 1} \right)\log t}}\left( {\rho \left| \sigma  \right.} \right) = J\left( {\rho \left| \sigma  \right.} \right)$, since $\left( {{\Delta _{\sigma ,\rho }}\log {\Delta _{\sigma ,\rho }}} \right)\left( Y \right) = \sigma \log \sigma \left( Y \right){\rho ^{ - 1}} - \sigma {\rho ^{ - 1}}\log \rho \left( Y \right)$.

Finally, we show ${\left\| {\rho  - \sigma } \right\|_1} = {\left\| {P - Q} \right\|_1}$.  Let $\fA = C^*(\rho_1-\rho_2)$ be commutative $C^*$-algebra generated by $\rho_1-\rho_2$, $M_n$ be the set of all $n\times n$ matrices and set the map $\fE:M_n \to \fA$ as trace preserving, conditional expectation. If we take $p_1 = \fE(\rho_1)$ and  $p_2 = \fE(\rho_2)$, then two elements
${\left( {{\rho _1} - {\rho _2}} \right)_ + }$ and ${\left( {{\rho _1} - {\rho _2}} \right)_ - }$ of Jordan decomposition of $\rho_1-\rho_2$, are commutative functional calculus
of $\rho_1-\rho_2$, and we have
${p_1} - {p_2} = \fE\left( {{\rho _1} - {\rho _2}} \right) = \fE\left( {{{\left( {{\rho _1} - {\rho _2}} \right)}_ + } - {{\left( {{\rho _1} - {\rho _2}} \right)}_ - }} \right)\, = \fE\left( {{{\left( {{\rho _1} - {\rho _2}} \right)}_ + }} \right) - \fE\left( {{{\left( {{\rho _1} - {\rho _2}} \right)}_ - }} \right) = {\left( {{\rho _1} - {\rho _2}} \right)_ + } - {\left( {{\rho _1} - {\rho _2}} \right)_ - }\, = {\rho _1} - {\rho _2}$ which implies ${\left\| {\rho  - \sigma } \right\|_1} = {\left\| {P - Q} \right\|_1}$.

\hfill \qed

\section{ACKNOWLEDGMENTS}
The author (S. F.) was partially supported by JSPS KAKENHI Grant Number 16K05257. 



\section{Appendix: Added notes related to Theorem 1}

Actually we have $\lim_{q\to 1} {\overline n_q} = \sum_{u\in\fU} p(u) l(u)$ which is the usual average code length,  but the definition of ${\overline n_q}$ in Theorem 1 seems to be complicated and somewhat unnatural to understand its meaning.  In order to overcome this problem, we may adopt the simple alternative definition for ${\overline n_q}$ instead of that in Theorem 1. Then we have the following proposition.
\\

{\bf Proposition A}
Let $q \geq 1$ and $c_{d,l,q} \leq 1$. Then we have

\[\frac{1}{2}\sum\limits_{u \in \fU} {\left| {p\left( u \right) - {Q_{d,l,q}}\left( u \right)} \right|}  \le \min \left\{ {1,\sqrt {\frac{{{\Delta _{d,q}}{{\log }_e}d}}{2}} } \right\}\]

Where ${\Delta _{d,q}} \equiv {\overline n _q} - {H_{d,q}}\left( \fU \right)$,${\overline n _q} \equiv \sum\limits_{u \in \fU } {p{{\left( u \right)}^q}l\left( u \right)}$,${H_{d,q}}\left( \fU \right) \equiv  - \frac{1}{{{{\log }_e}d}}\sum\limits_{u \in \fU} {p{{\left( u \right)}^q}{{\ln }_q}p\left( u \right)}$,
${Q_{d,l,q}}\left( u \right) \equiv \frac{1}{{{c_{d,l,q}}}}{\exp _q}\left( {{{\log }_e}{d^{ - l\left( u \right)}}} \right)$ and 
 ${c_{d,l,q}} \equiv \sum\limits_{u \in \fU} {{{\exp }_q}\left( {{{\log }_e}{d^{ - l\left( u \right)}}} \right)}$, where $q$-exponential function $\exp_q(\cdot)$ is the inverse function of $q$-logarithmic function $\ln_q(\cdot)$ and its form is given in the proof of this proposition. 

{\it Proof:}
By the same way to the proof of Theorem 1, we have 
\[{D_q}\left( {P\left\| {{Q_{d,l,q}}} \right.} \right) \ge \frac{1}{2}{\left( {\sum\limits_{u \in \fU} {\left| {p\left( u \right) - {Q_{d,l,q}}\left( u \right)} \right|} } \right)^2},\]
By simple computations with formula $\ln_q\frac{y}{x} =y^{1-q}(\ln_q y + \ln_q \frac{1}{x})$, we have
\begin{eqnarray*}
{D_q}\left( {P\left\| {{Q_{d,l,q}}} \right.} \right)&=&\sum_{u\in \fU} p(u)^q \left(\ln_q p(u) - \ln_q Q_{d,l,q}(u)\right) =-\log_e d \cdot H_{d,q}(\fU) -\sum_{u\in\fU} p(u)^q \ln_q \frac{\exp_q\left(\log_e d^{-l(u)}\right)}{c_{d,l,q}}  \\
&=&-\log_e d \cdot H_{d,q}(\fU) -\sum_{u\in \fU} p(u)^q \left( \exp_q\left(\log_e d^{-l(u)} \right)\right)^{1-q} \ln_q\frac{1}{c_{d,l,q}} -\sum_{u \in \fU} p(u)^q \log_e d^{-l(u)} \\
&=&\log_e d \cdot \sum_{u \in \fU} p(u)^q l(u)  -\log_e d \cdot H_{d,q}(\fU) -\sum_{u\in \fU} p(u)^q \left( \exp_q\left(\log_e d^{-l(u)} \right)\right)^{1-q} \ln_q\frac{1}{c_{d,l,q}} \\
&=&\Delta _{d,q}\log_e d  -\sum_{u\in \fU} p(u)^q \left( \exp_q\left(\log_e d^{-l(u)} \right)\right)^{1-q} \ln_q\frac{1}{c_{d,l,q}} \\
&\le& {\Delta _{d,q}} {\log _e}d 
\end{eqnarray*}
since  $d \geq 2$, $l(u) \geq 1$ implies $\log_e d^{-l(u)} \leq 0$ thus we have
$1+(1-q) \log_e d^{-l(u)} \geq 0$, 
then the definition of $q$-exponential function 
\[{\exp _q}\left( x \right) = \left\{ \begin{array}{l}
{\left( {1 + \left( {1 - q} \right)x} \right)^{\frac{1}{{1 - q}}}}\,,{\rm{if}}\,\,1 + \left( {1 - q} \right)x > 0\\
\,\,\,\,\,0\,\,\,\,\,\,\,\,\,\,\,\,\,\,\,\,\,\,\,\,\,\,\,\,\,\,\,\,\,\,\,\,\,,{\rm{otherwise}}
\end{array} \right.\]
shows $\exp_q(\log_e d^{-l(u)}) \geq 0$ and $c_{d,l,q} \leq 1$ was used.
Thus we have $\frac{1}{2}\left( \sum\limits_{u \in \fU} {\left| {p\left( u \right) - {Q_{d,l,q}}\left( u \right)} \right|}  \right)^2 \le   {\Delta _{d,q}{\log _e}d}.$

\hfill \qed
\\

We could not 
remove the needless and meaningless condition $c_{d,l,q} \leq 1$ in the above proposition, unfortunately. It is known that the inequality $c_{d,l,1} \leq 1 $ holds for the uniquely decodable code and the equality $c_{d,l,1}=1$ holds if the code archives the entropy, namely ${\overline n_1} = H_{d,1}(\fU)$ \cite{Sas2015}. 
In our proposition, we obtained $q$-parametric extension but it does not have any information theoretical meaning. We will have to consider about this problem in the future.

\end{document}